\documentclass[%
 reprint,
 amsmath,amssymb,
 aps,
]{revtex4-2}

\usepackage{graphicx}
\usepackage{dcolumn}
\usepackage{bm}
\usepackage{bbm}
\usepackage{subcaption}
\usepackage{comment}

\begin{document}

\preprint{APS/123-QED}

\title{Continuous-variable entanglement in a two-mode lossy cavity: an exact solution}
\author{Colin Vendromin}
\email{colin.vendromin@queensu.ca}
\author{Marc M. Dignam}
\affiliation{%
Department of Physics, Engineering Physics, and Astronomy, Queen's University, Kingston, Ontario K7L 3N6, Canada
}%

\date{\today}
\begin{abstract}
Continuous-variable (CV) entanglement is a valuable resource in the field of quantum information. One source of CV entanglement is the correlations between the position and momentum of photons in a two-mode squeezed state of light. In this paper, we theoretically study the generation of squeezed states, via spontaneous parametric downconversion (SPDC), inside a two-mode lossy cavity that is pumped with a classical optical pulse. The dynamics of the density operator in the cavity is modelled using the Lindblad master equation, and we show that the exact solution to this model is the density operator for a two-mode squeezed thermal state, with a time-dependent squeezing amplitude and average thermal photon number for each mode. We derive an expression for the maximum entanglement inside the cavity that depends crucially on the difference in the losses between the two modes. We apply our exact solution to the important example of a microring resonator that is pumped with a Gaussian pulse. The expressions that we derive will help researchers optimize CV entanglement in lossy cavities. 
\end{abstract}

\maketitle

\section{Introduction}
Entanglement serves as a basis for many applications in the field of quantum information \cite{Braunstein2005QuantumVariables}, such as quantum teleportation \cite{Metcalf2014QuantumChip, Bennett1992TeleportingChannels} and quantum key distribution \cite{HilleryQuantumStates, Jennewein2000QuantumPhotons}. Entanglement can occur as correlations between discrete variables of two particles, such as the polarization of two photons, or as the correlations between continuous variables, such as position and momentum of photons. In quantum-optical approaches to quantum computing, those that use discrete-variable (DV) entanglement can achieve high-fidelity operations, but their operation is probabilistic. The use of continuous-variable (CV) entanglement has the advantage that it can achieve deterministic operations, but is not able to achieve as high-fidelity operations \cite{Masada2015Continuous-variableChip}. To remedy this trade-off and attempt to achieve large-scale quantum computing, hybrid approaches that utilize both CV and DV have been demonstrated \cite{Takeda2019TowardComputing}.

In the context of quantum optics, two-mode squeezed states are routinely used as a source of CV entanglement. They can be generated by mixing two single-mode squeezed states on a beam-splitter \cite{Masada2015Continuous-variableChip}, or via a nonlinear interaction \cite{Villar2005GenerationEntanglement} such as spontaneous parametric downconversion (SPDC), where a strong coherent pump field interacts with a material that has a $\chi^{(2)}$ nonlinearity. The resulting two beams of squeezed light are entangled by the correlations between the position and momentum of the photons in each beam.

The amount of CV entanglement in the squeezed state increases with the amount of squeezing, so in the limit of infinite squeezing the two-mode squeezed state resembles a maximally-entangled EPR state \cite{Reid1989DemonstrationAmplification}. To enhance the amount of squeezing, and thus entanglement, it is beneficial for the SPDC interaction to occur in a cavity that is resonant with the pump field, as well as the generated signal and idler fields. 

In this work, we model the evolution of a squeezed state inside a lossy multimode cavity. We model the evolution with the Lindblad master equation. We show that the exact solution to this model is a two-mode squeezed thermal state for all time, with time-dependent squeezing amplitude, and thermal photon number for each mode.

In previous work \cite{Seifoory2017SqueezedCavities} we studied single-mode squeezed states in a lossy cavity. We modelled the dynamics of the generated squeezed light using the Lindblad master equation for a single-mode lossy cavity and derived the exact solution for the state in the cavity to be a squeezed thermal state. This present work builds on our previous work by including two modes that have different frequencies and loss rates. As one of the main results in this paper, we derive an analytic expression for the entanglement in the cavity as a function of the time-dependent squeezing and mode thermal photon numbers. This expression includes an explicit dependence on the mode losses and shows that the entanglement is maximized when the losses of the two modes are equal. 

An important structure for generating CV entanglement is a microring resonator side-coupled to a waveguide. It enhances the nonlinear interaction that produces squeezed light and can be integrated on a chip \cite{Yang2007EnhancedResonator, Guo2017ParametricChip,Liu2018Ultra-high-QPlatform,Silverstone2015QubitChip}. In the second part of the paper, we apply our model to a side-coupled ring resonator that is pumped with a Gaussian pulse that is coupled in from the linear waveguide. We derive a semi-analytic expression for the maximum entanglement in the ring that depends on the pump duration, coupling and scattering loss in the ring. 

The paper is organized as follows. In Sec. \ref{solution} we derive the two-mode squeezed thermal state solution, and give a set of coupled first-order differential equations for the squeezing amplitude and thermal photon numbers of each mode. Using our exact solution for the state in the cavity, in Sec. \ref{cv} we derive an expression for the inseparability of the state, and the maximum entanglement in the cavity as a function of the input pump amplitude. In Sec. \ref{example} we apply our theory to the example of generating CV entanglement in a side-coupled microring resonator pumped with a Gaussian input pulse. Finally in Sec. \ref{conclusion} we present our conclusions. 

\section{The Lindblad master equation for a two-mode lossy cavity}
\label{solution}
In this section we derive the exact solution to the Lindblad master equation for the generation of a two-mode squeezed state in a lossy cavity. The cavity is pumped with a strong pump field, and signal and idler photon pairs are generated via spontaneous parametric downconversion (SPDC) with frequencies $\omega_1$ and $\omega_2$, respectively. We assume that the cavity is resonant at the  frequencies of the signal and idler. Treating the pump as a strong classical field and neglecting depletion of the pump due to the nonlinear interaction, the Hamiltonian for the signal and idler photons in the cavity is given by \cite{quantumopticsGarrison}
\begin{equation}
    \label{hamiltonian}
    \hat{H} = \hbar\omega_1 \hat{b}_1^\dagger\hat{b}_1+ \hbar\omega_2 \hat{b}_2^\dagger\hat{b}_2 + \gamma^* \mathcal{E}^*_P(t)  \hat{b}_1\hat{b}_2+ \gamma\mathcal{E}_P(t) \hat{b}_1^\dagger\hat{b}_2^\dagger,
\end{equation}
where, $\hat{b}_1$ ($\hat{b}_1^\dagger$) and $\hat{b}_2$ ($\hat{b}_2^\dagger$) destroy (create) photons in the cavity in the mode 1 and 2 with frequency $\omega_1$ and $\omega_2$, respectively. The last two terms in Eq. \eqref{hamiltonian} account for the nonlinear SPDC interaction that generates the two-mode squeezed light, where the second-order nonlinear coefficient is $\gamma = -i\hbar \chi^{(2)}\sqrt{\omega_1 \omega_2}$, where $\chi^{(2)}$ is an effective second-order nonlinear susceptibility. Here $\mathcal{E}_P(t) $ is the positive frequency part of the time-dependent classical electric field, $E_P(t)$, where $E_P(t) = \mathcal{E}_P(t) + \mathcal{E}^*_P(t)$. In all that follows, only the positive frequency part of the electric field is used, because we make the rotating wave approximation.

To gain some insight into the state of the light inside the cavity, first consider the case that the cavity is lossless and the pump is a continuous wave with $\mathcal{E}_P(t) = \mathcal{E}_0 \exp(-i(\omega_1+\omega_2) t)$. Then the state evolves in time with a unitary two-mode squeezing operator \cite{quantumopticsGarrison} given by
\begin{equation}
    \label{squeezingop}
    \hat{S}(\xi) = {\rm e}^{\xi^*\hat{b}_1\hat{b}_2 - \xi\hat{b}_1^\dagger\hat{b}_2^\dagger },
\end{equation}
with a time-dependent complex squeezing parameter $\xi = u \exp (i\phi)$, related to the pump amplitude by $\xi = it\gamma \mathcal{E}_0/\hbar$. 

Now we consider the general case where the cavity has loss and there is a pulsed pump. In this case, the dynamics of the density operator $\hat{\rho}$ in the cavity are modelled using the Lindblad master equation \cite{openQsystemsBreuer}:
\begin{equation}
    \label{lindblad}
    \frac{d\hat{\rho}}{dt} = -\frac{i}{\hbar}\left[\hat{H},\hat{\rho}\right]+\frac{1}{2}\sum_{j=1}^2\Gamma_j\left(2\hat{b}_j\hat{\rho}\hat{b}_j^\dagger - \hat{b}_j^\dagger\hat{b}_j\hat{\rho}- \hat{\rho}\hat{b}_j^\dagger\hat{b}_j\right),
\end{equation}
where the loss of the cavity is captured by intensity decay rates $\Gamma_1$ and $\Gamma_2$ for modes 1 and 2, respectively. With the inclusion of loss, the state does not simply evolve in time by operating with a squeezing operator. 

The main result of this paper is the derivation of the exact solution to the system of Eqs. \eqref{hamiltonian} and \eqref{lindblad}. First, however, we consider a simpler situation where there is no pump in the cavity ($\mathcal{E}_P(t) = 0$), so that the last two terms in the Hamiltonian in Eq. \eqref{hamiltonian} are zero. In this case, we assume that the initial state is a two-mode thermal state, which is formed by the product of single-mode thermal states:
\begin{eqnarray}
    \hat{\rho}_{th} = \prod_{j=1}^{2} \left(1-x_j\right) \left(x_j\right)^{\hat{n}_j},
    \label{2modethermalstate}
\end{eqnarray}
where
\begin{equation}
\label{boltzmannfactor}
    x_j = \exp\left(-\frac{\hbar \omega_j}{k_B T_j}\right)
\end{equation}
and $\hat{n}_j$ is the photon number operator for the mode $j$.
Here, $k_B$ is the Boltzmann constant and  $T_j$ is the effective temperature of mode $j$. The average number of photons in the thermal state in each mode $n_j$ is given by
\begin{equation}
    \label{thermalphotonnumber}
    n_j = \frac{x_j}{1-x_j}.
\end{equation}
We now prove that the state in the cavity remains a two-mode thermal state for all time, with an average thermal photon number that decays exponentially over time. This means that the solution to Eq. \eqref{lindblad} when there is no interaction term in the Hamiltonian in Eq. \eqref{hamiltonian} can be written as
\begin{equation}
    \label{solution1}
    \hat{\rho}(t) = \hat{\rho}_{th}(t),
\end{equation}
where $\hat{\rho}_{th}(t)$ is given in Eq. \eqref{2modethermalstate}, except now the variable $x_j$ is time-dependent due to the dynamical thermal photon number. Rearranging Eq. \eqref{solution1} we obtain
\begin{equation}
\label{constantofmotion1}
   \hat{\mathbbm{1}} = \left[\hat{\rho}_{th}(t)\right]^{-1/2}\hat{\rho}(t)\left[\hat{\rho}_{th}(t)\right]^{-1/2},
\end{equation}
where $\hat{\mathbbm{1}}$ is the identity operator and we have used the fact that the thermal state operator is unitary.  Taking the time derivative of both sides of Eq. \eqref{constantofmotion1} yields

\begin{equation}
    0 = \frac{d}{dt}\left(\hat{\rho}_{th}^{-1/2}\hat{\rho}\hat{\rho}_{th}^{-1/2}\right).
    \label{timederivative1}
\end{equation}
Applying the chain rule, we obtain
\begin{eqnarray}
\label{timederivative1_chainrule}
    0&=&\frac{d \hat{\rho}_{th}^{-1/2}}{dt}\hat{\rho}\hat{\rho}_{th}^{-1/2} + \hat{\rho}_{th}^{-1/2}\frac{d\hat{\rho}}{dt}\hat{\rho}_{th}^{-1/2} + \hat{\rho}_{th}^{-1/2}\hat{\rho}\frac{d \hat{\rho}_{th}^{-1/2}}{dt}. \nonumber
    \\
\end{eqnarray}
 Simplifying Eq. \eqref{timederivative1_chainrule} using the derivative of $\hat{\rho}$ in Eq. \eqref{lindblad} and the identity in Eq. \eqref{constantofmotion1}, we eliminate $\hat{\rho}$ from all the terms in Eq. \eqref{timederivative1_chainrule}, and are left with terms that only contain $\hat{\rho}_{th}$ and its derivative. Then, using the thermal state in Eq. \eqref{2modethermalstate}, we obtain 
\begin{eqnarray}
\label{characteristicEq1}
    0&=&\sum_{j=1}^2 \left(\hat{n}_j + \hat{\mathbbm{1}} \frac{x_j}{x_j-1}\right)D_j,
\end{eqnarray}

where
\begin{equation}
\label{infrontofn}
    D_j = -\frac{1}{x_j}\frac{dx_j}{dt}+\Gamma_j\left(x_j -1\right).
\end{equation}
In order for Eq. \eqref{characteristicEq1} to be true for all times we must have that $D_j = 0$, which has the solution

\begin{equation}
\label{xsolutionforthermalstate}
    x_j(t) = \frac{1}{1+\mathcal{C}{\rm e}^{\Gamma_j t}},
\end{equation}
where $\mathcal{C}$ is a constant determined by the initial conditions. Using Eqs. \eqref{boltzmannfactor} and \eqref{thermalphotonnumber}, we can rewrite Eq. \eqref{xsolutionforthermalstate} as
\begin{equation}
\label{nthsoltution}
    n_j(t) = n_j(0){\rm e}^{-\Gamma_j t},
\end{equation}
which says that the average photon number in each mode decays from an initial value $n_j(0)$ at the rate $\Gamma_j$. Thus, if we start with a thermal state, it will remain a thermal state for all times, but with different time-dependent temperatures for the two modes.

Our focus now is to include the pump, so that the last two terms in the Hamiltonian in Eq. \eqref{hamiltonian} are not zero.
In this case, as we shall see, the two-mode thermal state will be squeezed with the two-mode squeezing operator $\hat{S}$ in Eq. \eqref{squeezingop}. Therefore, we propose that the solution to Eq. \eqref{lindblad} is the squeezed two-mode thermal state
\begin{equation}
    \label{solution2}
    \hat{\rho}(t) = \hat{S}(\xi(t))\hat{\rho}_{th}(t)\hat{S}^\dagger(\xi(t)),
\end{equation}
where now the thermal photon number and squeezing parameter are time-dependent. Rearranging Eq. \eqref{solution2} we obtain
\begin{equation}
    \label{constantofmotion2}
    \hat{\mathbbm{1}} = \left[\hat{\rho}_{th}(t)\right]^{-1/2}\hat{S}^\dagger(\xi(t))\hat{\rho}\hat{S}(\xi(t))\left[\hat{\rho}_{th}(t)\right]^{-1/2}.
\end{equation}
Taking the time derivative of both sides, we obtain

\begin{equation}
\label{timederivative2}
   0 = \frac{d}{dt}\left(\hat{\rho}_{th}^{-1/2}\hat{S}^\dagger\hat{\rho}\hat{S}\hat{\rho}_{th}^{-1/2}\right).
\end{equation}
Applying the chain rule gives
\begin{align}
0 &=\frac{d\hat{\rho}_{th}^{-1/2}}{dt}\hat{S}^\dagger\hat{\rho}\hat{S}\hat{\rho}_{th}^{-1/2} +\hat{\rho}_{th}^{-1/2}\hat{S}^\dagger\hat{\rho}\hat{S}\frac{d\hat{\rho}_{th}^{-1/2}}{dt} + \nonumber
\\
&+\hat{\rho}_{th}^{-1/2}\frac{d\hat{S}^\dagger}{dt}\hat{\rho}\hat{S}\hat{\rho}_{th}^{-1/2} + \hat{\rho}_{th}^{-1/2}\hat{S}^\dagger\hat{\rho}\frac{d\hat{S}}{dt}\hat{\rho}_{th}^{-1/2} + \label{timederivative2_terms}
\\
&+\hat{\rho}_{th}^{-1/2}\hat{S}^\dagger\frac{d\hat{\rho}}{dt}\hat{S}\hat{\rho}_{th}^{-1/2}. \nonumber
\end{align}
Following a process similar to the one above in the case of no pump, in Appendix \ref{appndx1} we simplify the terms in Eq. \eqref{timederivative2_terms} and show that for the equality to be satisfied for all times, the thermal photon numbers ($n_1(t)$ and $n_2(t)$), squeezing amplitude ($u(t)$), and squeezing phase ($\phi(t)$) must be solutions of the following first-order coupled differential equations:
\begin{align}\label{nth1_deq}
    \frac{dn_1}{dt} &= n_1\left(\Gamma_2\sinh^2{u} - \Gamma_1\cosh^2{u}\right)+\Gamma_2\sinh^2{u},
    \\
    \label{nth2_deq}
    \frac{dn_2}{dt} &= n_2\left(\Gamma_1\sinh^2{u} - \Gamma_2\cosh^2{u}\right)+\Gamma_1\sinh^2{u},
    \\
    \label{u_deq}
    \frac{d u}{dt} &= \frac{i}{2\hbar}\left(\mathcal{E}_P\gamma{\rm e}^{-i\phi}-\mathcal{E}^*_P\gamma^*{\rm e}^{i\phi}\right) -\frac{\sinh(2u)}{n_1+n_2+1}\nonumber
    \\
    &\times\left(\frac{\Gamma_1 + \Gamma_2}{2} + \frac{\Gamma_1 - \Gamma_2}{2}\left[n_2 - n_1\right] \right),
    \\
    \label{phi_deq}
    \frac{d \phi}{dt} &= -\left(\omega_1+\omega_2\right) +\frac{\left(\mathcal{E}_P\gamma{\rm e}^{-i\phi}+\mathcal{E}^*_P\gamma^*{\rm e}^{i\phi}\right)}{\hbar\tanh(2u)},
\end{align}
 where we have omitted the time-dependencies in all variables and $\mathcal{E}_P(t)$ for simplicity. Eqs. \eqref{nth1_deq} to \eqref{phi_deq} describe the dynamics of the squeezing of the state in the cavity. If we let the decay rates of the two modes be identical, $i.e.\, \Gamma_1 = \Gamma_2$,  this set of equations reduces to the single mode squeezing equations from our previous work \cite{Vendromin2020OptimizationStates}. Note that Eq. \eqref{phi_deq} only depends on the sum-frequency frequency $\omega_s  = \omega_1 + \omega_2$. Thus, the dynamics of the squeezed state does not depend on the individual frequencies of the signal and idler photons, except implicitly through the effective nonlinearity, $\gamma$. 
 
 In order to obtain a specific solution to this set of first-order differential equations, an initial state in the cavity must be specified. In all that follows, we take the initial state in the cavity to be the vacuum. Other initial states can be chosen, but we chose the vacuum state because, as we will show, it simplifies the form of the equations and corresponds to physically-realizable experiments. 
 
 We define an initial time $t_i$ when the state in the cavity is vacuum, that is $u(t_i) = 0$, and $n_j(t_i) = 0$, respectively. This causes the denominator of the second term in Eq. \eqref{phi_deq} to be zero, because $\tanh(2 u(t_i)) = 0$. In order to eliminate the singularity, we choose the initial squeezing phase such that the numerator of the second term is zero
\begin{equation}
\label{remove_singularity}
   \mathcal{E}_P(t_i)\gamma{\rm e}^{-i\phi(t_i)}+\mathcal{E}^*_P(t_i)\gamma^*{\rm e}^{i\phi(t_i)} = 0,
\end{equation}
where the initial pump in the cavity $\mathcal{E}_P(t_i)$ is arbitrarily small but not zero. If we let the pump and nonlinear parameter be written as a general complex number as $\mathcal{E}_P(t_i)\gamma = |\mathcal{E}_P(t_i)||\gamma|{\rm e}^{i\theta-i\omega_st_i}$, then the solution to Eq. \eqref{remove_singularity} for the initial squeezing phase is $\phi(t_i) =  - \pi/2 - \theta + \omega_s t_i$ \cite{initial_phase_note}. Iterating Eqs. \eqref{u_deq} and \eqref{phi_deq} forward one step in time from the initial condition, it can be shown that the squeezing phase for all future times is given by
\begin{align}
    \label{phi_deq_final}
    \phi(t) &= -\omega_s t +\phi(t_i)\nonumber
    \\
    &=-\omega_s\left(t-t_i\right) -\frac{\pi}{2}+\theta.
\end{align}
 Therefore if the initial state is vacuum, then the squeezing phase does not depend on the thermal photon numbers or squeezing amplitude, and simply rotates around the origin of phase space with the sum-frequency $\omega_s$.

Using the expression in Eq. \eqref{phi_deq_final} for $\phi(t)$ in Eq. \eqref{u_deq}, the equation for the squeezing amplitude becomes, 
\begin{align}
    \label{u_deq_simp}
     \frac{d u}{dt} &= \frac{|\mathcal{E}_P(t)||\gamma|}{\hbar} - \frac{\sinh(2u)}{n_1+n_2+1}\nonumber
    \\
    &\times\left(\frac{\Gamma_1 + \Gamma_2}{2} + \frac{\Gamma_1 - \Gamma_2}{2}\left[n_2 - n_1\right] \right).
\end{align}
It is convenient to write Eqs. \eqref{nth1_deq}, \eqref{nth2_deq}, and \eqref{u_deq_simp} in terms of the dimensionless parameter $\tilde{t} = \Gamma_{+} t$, where $\Gamma_{\pm} = (\Gamma_1 \pm \Gamma_2)/2$.  Doing this, we obtain
\begin{align}
    \label{u_deq_final}
     \frac{d u}{d\tilde{t}} &= \frac{g(\tilde{t})}{2} - \frac{\sinh(2u)}{n_1+n_2+1}\left(1+ \zeta\left[n_2 - n_1\right] \right),
     \\
     \label{nth1_deq_final}
    \frac{dn_1}{d\tilde{t}} &= n_1\left(\left[1-\zeta\right]\sinh^2{u} - \left[1+\zeta\right]\cosh^2{u}\right)+ \nonumber
    \\
    &+\left[1-\zeta\right]\sinh^2{u},
    \\
    \label{nth2_deq_final}
    \frac{dn_2}{d\tilde{t}} &= n_2\left(\left[1+\zeta\right]\sinh^2{u} - \left[1-\zeta\right]\cosh^2{u}\right)+ \nonumber
    \\
    &+\left[1+\zeta\right]\sinh^2{u},
\end{align}
where $\zeta=\Gamma_{-}/\Gamma_{+}$ is proportional to the difference of the two decay rates, and $g(t)$ is the pumping strength in the ring; it is defined as the rate of light generation in the cavity divided by the average decay rate of light out of the cavity
\begin{equation}
    \label{pumpingstrengthdef}
    g(t) \equiv \frac{2|\mathcal{E}_P(t)||\gamma|}{\hbar \Gamma_{+}}.
\end{equation}
 Because the decay rates must be positive, it follows that $|\zeta| < 1$. When $\zeta = 0$, the decay rates of the modes are equal, the thermal photon numbers are the same, and consequently the coupled equations reduce to the single-mode squeezing case. 

In summary, we have shown that the exact solution to the Lindblad master equation for the generation of a two-mode squeezed state in a lossy cavity is a two-mode squeezed thermal state and we have derived a set of coupled first-order differential equations that describe the dynamics of the state as a function of the pumping strength and cavity decay rates for each mode. We now use the two-mode squeezed thermal state to give a condition for the continuous variable entanglement in the cavity, and derive an expression for the maximum entanglement. 

\section{Continuous-variable entanglement in a two-mode squeezed thermal state}
\label{cv}
In this section we give the entanglement condition for the light in the cavity and derive a semi-analytic expression for the maximum amount of entanglement as a function of the pump and the thermal photon number difference between the two modes. 

We define a general quadrature operator
\begin{equation}
\label{gen_quad}
    \hat{\chi}_j(\beta_j) = \frac{\hat{b}_j{\rm e}^{i\beta_j} + {\hat{b}}^\dagger_j{\rm e}^{-i\beta_j}}{2},
\end{equation}
where $j=1$ and $j=2$ are the two modes in the cavity, and $\beta_j$ is an angle in phase space. The quantum noise in the quadrature is given by
\begin{equation}
\label{gen_quad_noise}
    \Delta \chi_j =\sqrt{ {\rm tr}\left(\hat{\rho}\hat{\chi}_j^2\right) -  \left[{\rm tr}\left(\hat{\rho}\hat{\chi}_j\right)\right]^2 }.
\end{equation}
For the two-mode squeezed thermal state in Eq. \eqref{constantofmotion2}, using the fact that ${\rm tr}(\hat{\rho}\hat{\chi}_j)  = 0$,  we obtain for the quadrature noise
\begin{equation}
    \label{gen_quad_noise_2STS}
   \Delta \chi_j= \sqrt{\frac{1}{4}\cosh(2u)+\frac{1}{2}\cosh^2(u)n_j+\frac{1}{2}\sinh^2(u)n_k},
\end{equation}
where $k=1$ if $j=2$ and $k=2$ if $j=1$. The vacuum noise in the quadrature is given by $\Delta \chi_j = 1/2$. Therefore the quadrature noise is squeezed below the vacuum noise whenever $\Delta \chi_j < 1/2$. For the two-mode squeezed thermal state, the quadrature noise in Eq. \eqref{gen_quad_noise_2STS} is represented as a circle in phase space with a radius always greater than or equal to $1/2$, so there is no squeezing. Linear combinations of the quadrature operators for mode $1$ and $2$, however, can exhibit squeezing. To this end, we define two operators $\hat{X}$ and $\hat{Y}$ 
\begin{align}
    \label{X_sum}
    \hat{X} &\equiv \hat{\chi}_1\left(\beta_1\right) + \hat{\chi}_2\left(\beta_2\right)
    \\
    \label{Y_diff}
   \hat{Y}&\equiv \hat{\chi}_1\left(\beta_1 +\frac{\pi}{2}\right) - \hat{\chi}_2\left(\beta_2+\frac{\pi}{2}\right). 
\end{align}
The operators $\hat{X}$ and $\hat{Y}$ are proportional to the sum and difference of the individual position and momenta operators of the light in each mode. The quadrature noise in $\hat{X}$ and $\hat{Y}$ using the two-mode squeezed thermal state is
\begin{align}
    \label{noiseinX}
    \left<(\Delta X)^2\right> &= \frac{1}{2}\left(1+n_1+n_2\right)\nonumber
    \\
    &\times\left[\cosh(2u)-\cos\left(\phi+\beta_1+\beta_2\right)\sinh(2u)\right],
\end{align}
where $\phi$ is the time-dependent squeezing phase in Eq. \eqref{phi_deq_final}. It is straightforward to show that $\left<(\Delta Y)^2\right> = \left<(\Delta X)^2\right> $. For certain phase relationships between $\beta_1(t)$, $\beta_2(t)$, and $\phi(t)$, the quadrature noise is reduced below vacuum. Choosing the relationship between the quadrature phases be $\beta_1(t) +\beta_2(t) = - \phi(t)$, the fast oscillations are removed and the quadrature noise is exponentially squeezed to give
\begin{align}
    \label{squeezing}
    \left<(\Delta X)^2\right> &= \frac{1}{2}\left[1+n_1(t)+n_2(t)\right]{\rm e}^{-2u(t)}.
\end{align}
In the limit of infinite squeezing, $u\rightarrow\infty$, the uncertainty in the relative quadrature operator goes to zero, $\left<(\Delta X)^2\right>\rightarrow0$. Thus, a measurement on $\hat{\chi}_1$ would give an exact prediction of $\hat{\chi}_2$, such that the two modes are perfectly correlated. However, for finite squeezing the two modes will not be perfectly correlated. To evaluate the degree of entanglement between the two modes, we define the correlation variance as the sum of the quadrature noises in the relative position and momentum operators
\begin{equation}
    \label{correlationfunc_gen}
   \Delta^2_{1,2} =  \left<(\Delta X)^2\right> +\left<(\Delta Y)^2\right>, 
\end{equation}
and using Eq. \eqref{squeezing} we obtain
\begin{equation}
\label{correlationfunc}
    \Delta^2_{1,2}(t) = \left[1+ n_1(t)+n_2(t)\right]{\rm e}^{-2u(t)}.
\end{equation}
It has been proved by Duan \textit{et al.} \cite{Duan2000InseparabilitySystems} and Simon \cite{Simon2000Peres-HorodeckiSystems} that the two-mode squeezed state is inseparable, and therefore entangled, when the correlation variance is less than 1,
\begin{equation}
    \label{entanglement_cond}
    \Delta^2_{1,2}(t)<1.
\end{equation}
 This condition is achieved when the exponential squeezing factor ($\exp(-2u)$) overcomes the thermal noise factor in front. 

The correlation variance in Eq. \eqref{correlationfunc} is time dependent because the squeezing amplitude and thermal photon numbers are time dependent. If we assume that the pump is pulsed, then there should be a time when the correlation variance is minimized and the state exhibits the maximum amount of entanglement. In order to find an expression for the maximum entanglement we find the extreme point of the correlation variance in Eq. \eqref{correlationfunc}, $i.e.$ we solve the following equation:
\begin{equation}
\label{solve_max_entangle}
    \frac{d \Delta_{1,2}^2}{dt}\bigg|_{t=t_{min}} = 0.
\end{equation}
Simplifying this equation by using Eqs. \eqref{u_deq_final} - \eqref{nth2_deq_final} we obtain,
\begin{equation}
\label{max_entangle}
     \left(\Delta_{1,2}^2\right)_{min} = \frac{1+\zeta \left[n_2(t_{min})-n_1(t_{min})\right]}{1+g(t_{min})}.
\end{equation}
Thus, even a small difference in the decay rates of the two modes can cause an increase in the minimum value of the correlation variance and decrease the degree of entanglement. In situations where we can treat $|\zeta|$ as a small perturbation, we can neglect the second term in the numerator of Eq. \eqref{max_entangle} and write,
\begin{equation}
    \label{approx_max_entangle}
     \left(\Delta_{1,2}^2\right)_{min} \approx \frac{1}{1+g(t_{min})}.
\end{equation}
This represents an ideal case, when the two decay rates for the modes are the same and we can drop the $\zeta$ dependence. In this case, better entanglement is always achieved by increasing the pumping strength. We do not have an expression for the time when the correlation variance is minimum, $t_{min}$, so Eq. \eqref{max_entangle} and Eq. \eqref{approx_max_entangle} must still be evaluated numerically. However, if the pump field is a continuous wave, then the pumping strength $g(t)$ does not depend on time, and the pumping strength at $t_{min}$ in Eq. \eqref{approx_max_entangle} can be replaced with essentially the amplitude of the pump in the cavity.  When the pump field is pulsed then, as we will show in the example below, it is often a good approximation to replace $g(t_{min})$ with the peak value of the pump in the cavity, $g_{max}$. In the next section, we will discuss the results of the entanglement for a two-mode squeezed thermal state generated in a side-coupled ring resonator. 
\section{Example: generating continuous-variable entanglement in a ring resonator }
\label{example}
In this section we apply our formalism to a ring resonator, which is optimized to produce a maximally-entangled state, and we study the dynamics of the entanglement. In previous work \cite{Vendromin2020OptimizationStates}, we treated the same problem but for degenerate squeezing, where the signal and idler have the same frequency.  In this example, we will generalize our previous results to the case where the signal and idler are distinguishable by their frequency and loss. 

\begin{figure}
    \centering
    \includegraphics[scale=0.333]{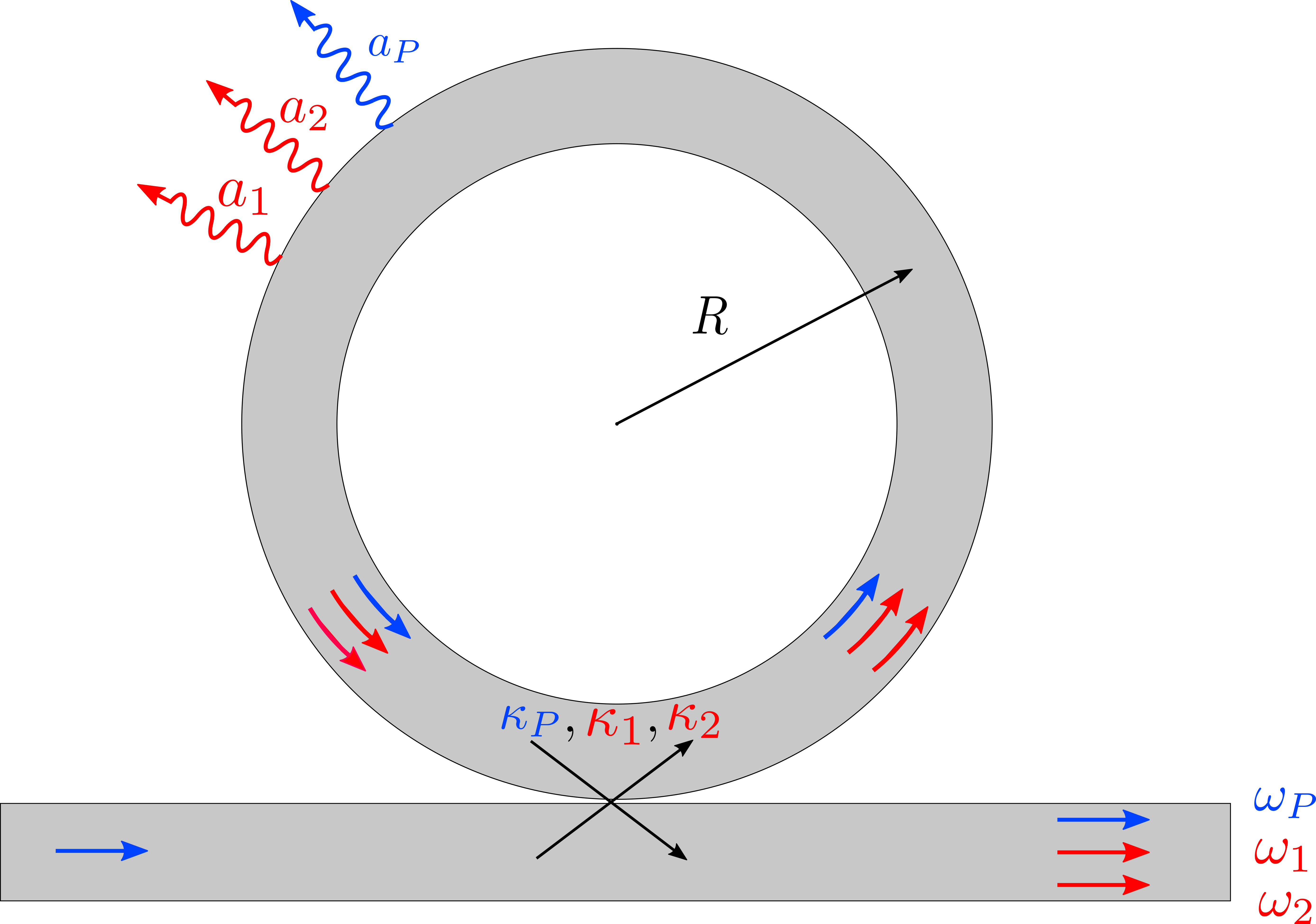}
    \caption{Schematic of the ring resonator side-coupled to a channel waveguide. The thick arrows represent the components of the pump field (blue) with frequency $\omega_P$, incident from the left, and the signal and idler fields (both red) with frequencies $\omega_1$ and $\omega_2$, respectively. The cross-coupling coefficients and scattering loss parameters for the pump, signal, and idler fields are $\kappa_P$, $\kappa_1$, and $\kappa_2$ and $a_P$, $a_1$, and $a_2$, respectively.}
    \label{fig:ringreso}
\end{figure}

The ring resonator system we consider, shown in Fig. \ref{fig:ringreso}, consists of a straight waveguide (channel) side-coupled to a ring waveguide with radius $R$. Both the ring and channel are made from the same material that has a nonlinear $\chi^{(2)}$ response, but we ignore that nonlinear interaction in the channel, since the pump field is much weaker there than in the ring. The classical pump field incident on the ring is taken to be a classical Gaussian pulse. The positive frequency part of the pump field in the channel takes the form
\begin{equation}
\label{ElectricFieldinCH}
    \mathcal{E}_{CH}(t) = \mathcal{E}_0\sqrt{\frac{T_R}{\tau}}\exp\left(-2\ln(2)\frac{t^2}{\tau^2}\right)\exp(-i\omega_Pt),
\end{equation}
where $\mathcal{E}_0$ is the amplitude, $\tau$ is the FWHM of the intensity of the Gaussian, $\omega_P$ is the central frequency, and $T_R=2\pi R n_{eff}/c$ is the ring round-trip time, where $n_{eff}$ is the effective index for the pump. We scale the amplitude by $1/\sqrt{\tau}$ so that the pulse energy is a constant that is independent of its duration. The coupling of the pulse into the ring happens at a single coupling point, as indicated in Fig. \ref{fig:ringreso}. The pump field in the ring, $\mathcal{E}_R(\omega)$, is calculated using a transfer matrix approach \cite{Vendromin2020OptimizationStates, 2008OpticalMicroresonators}, and is related to the field in the channel by
\begin{equation}
    \label{ringpump}
    \mathcal{E}_R(\omega) = \frac{i\kappa_P a_P\,{\rm e}^{i\omega T_R}}{1-\sigma_P a_P \,{\rm e}^{i\omega T_R} }\mathcal{E}_{CH}(\omega),
\end{equation}
where $\mathcal{E}(\omega)$ is the Fourier transform of $\mathcal{E}(t)$ defined as
\begin{equation}
\label{FT}
    \mathcal{E}(\omega) = \int_{-\infty}^{\infty}\mathcal{E}(t)\,{\rm e}^{i\omega t} dt,
\end{equation}
and the inverse Fourier transform as
\begin{equation}
\label{invFT}
    \mathcal{E}(t) = \frac{1}{2\pi}\int_{-\infty}^{\infty}\mathcal{E}(\omega)\,{\rm e}^{-i\omega t} d\omega.
\end{equation}
In Eq. \eqref{ringpump}, the constant $\kappa_P$ is the frequency-independent cross-coupling constant between the ring and channel, and is related to the self-coupling constant $\sigma_P$ through the lossless coupling relation: $\kappa_P^2+\sigma_P^2 = 1$. The parameter $a_P$ is the amplitude attenuation for the pump in the ring after a single round-trip. It is related to the power attenuation coefficient $\alpha_P$ by $a_P^2 = \exp(-\alpha_P 2\pi R)$. This is the power lost due to scattering only and not coupling. When $a_P=1$, there is no scattering loss. 

The buildup factor is defined as the ratio of the intensity in the ring to the channel and is given by
\begin{align}
    \label{buildup}
    \mathcal{B}(\omega)\equiv\frac{|\mathcal{E}_{R}(\omega)|^2}{|\mathcal{E}_{CH}(\omega)|^2} = \frac{\kappa_P^2 a_P^2}{1-2\sigma_P a_P \cos(\omega T_R) + \sigma_P^2 a_P^2}.
\end{align}
It contains resonant peaks when the incident light is on resonance with the ring, $i.e.$ $\cos(\omega T_R) = 1$, and the buildup is maximized. We choose the central frequency of the pump pulse to be on resonance with the ring, $\omega_P = 2\pi m_P/T_R$, where $m_P$ is a positive integer defining the pump mode number. Inside the ring, the pump field generates signal and idler photons at the frequencies $\omega_1$ and $\omega_2$, via SPDC, such that energy is conserved: $\omega_P = \omega_1+ \omega_2$. The signal and idler fields are also resonant with the ring, $\omega_1 T_R = 2\pi m_1$ and $\omega_2 T_R = 2\pi m_2$, where $m_1$ and $m_2$ are the mode numbers for the signal and idler fields, respectively. For simplicity we assume perfect phase matching between the three fields, and that the effective index of refraction is the same for the pump, signal, and idler modes. The latter assumption is experimentally realized in an AlN microring resonator with a waveguide width of $ 1.10{\rm \mu m}$ \cite{Guo2017ParametricChip}, and makes the ring round-trip time the same for each field. Using these assumptions the mode numbers are related by $m_P = m_1+ m_2$. 

Generally, signal and idler photons will be generated in a number of different pairs of ring modes, as long as they satisfy energy conservation and phase matching. To simply our theory however, we assume that only the two ring modes $m_1$ and $m_2$ are perfectly phased matched, and squeezed light generation in all other modes is neglected. 

Since the pump is pulsed, in general it will couple into multiple ring modes depending on how wide its bandwidth is. To ensure that most of the pump light couples into a single mode $m_P$ we require that its duration is longer than the ring round-trip time ($\tau > T_R$). Doing so makes the pulse bandwidth overlap a single resonance peak of the buildup factor and thus the adjacent peaks in the buildup factor do not significantly couple any light into the ring. The buildup factor for the central pump frequency $\omega_P$ becomes,
\begin{align}
    \label{builduppump}
    \mathcal{B}(\omega_P) = \frac{\kappa_P^2 a_P^2}{(1-\sigma_P a_P)^2}.
\end{align}
In all that follows we are only interested in the limit that the buildup factor is large, which only occurs when $(1-\sigma_P a_P) \ll 1$, such that the denominator of Eq. \eqref{builduppump} is small. This limit produces the largest entanglement in the ring. 

Now we need to find an expression for the time-dependent pump field in the ring that we can use to solve for the dynamics of the squeezed state. This is done by taking the inverse Fourier transform of the field component in the ring. As we have shown in our previous work \cite{Vendromin2020OptimizationStates}, in the limit when $(1-\sigma_P a_P) \ll 1$ the integral can be approximated very well by
\begin{equation}
    \label{ElectricFieldinR}
    \mathcal{E}_R(t) = \frac{\tau \kappa_P a_P  \sqrt{\pi}{\rm e}^{z(t)^2}{\rm erfc}[z(t)]}{\sqrt{8\ln(2)}\,T_R}\mathcal{E}_{CH}(t),
\end{equation}
where
\begin{equation}
    \label{z}
   z(t) \equiv \frac{(1-\sigma_P a_P)\tau}{\sqrt{8\ln(2)}\,T_R} - \frac{\sqrt{8\ln(2)}\,t}{2\tau}.
\end{equation}

Now to model the evolution of the state in the ring we use the pump field in Eq. \eqref{ElectricFieldinR} in the coupled Eqs. \eqref{u_deq_final} to \eqref{nth2_deq_final}. The pumping strength in the ring defined in Eq. \eqref{pumpingstrengthdef} becomes
\begin{align}
\label{pumpstrengthinring1}
g(\tilde{t}\,)&= g_0 \frac{\kappa_P a_P  \sqrt{\pi} {\rm e}^{z(\tilde{t}\,)^2}{\rm erfc}[z(\tilde{t}\,)]}{\sqrt{8\ln(2)}}\sqrt{\frac{\tilde{\tau}}{T_R\Gamma_+}}\nonumber
\\
&\times \exp\left(-2\ln(2)\frac{{\tilde{t}}^2}{{\tilde{\tau}}^2}\right),
\end{align}
where we define the dimensionless parameter 
\begin{equation}
\label{g0}
g_0 \equiv \frac{2|\gamma|\mathcal{E}_0}{\hbar \Gamma_+}, 
\end{equation}
which is proportional to the input pump pulse amplitude $\mathcal{E}_0$ divided by the average decay rate of the signal and idler mode $\Gamma_+$, where we have used the same parameters defined above,  $\tilde{t} = \Gamma_+ t$ and $\tilde{\tau} = \Gamma_+ \tau$. At this point we could simply solve the dynamic Eqs. \eqref{u_deq_final} to \eqref{nth2_deq_final} numerically for any desired loss and pump parameters. However, as we shall show, a key parameter that affects the entanglement is the difference in the loss in the signal and idler modes, $\zeta$. Therefore, to simplify the discussion of our results and to make the pumping strength in Eq. \eqref{pumpstrengthinring1} dependent only on the pump coupling and loss parameters, we assume that the average of the decay rate of the signal and idler is equal to the pump decay rate, $\Gamma_P$, $i.e.$, $\Gamma_+ \equiv \Gamma_P$, where when $(1-\sigma_P a_P)\ll1$, the pump decay rate is
\begin{equation}
\label{pumpdecayrate}
    \Gamma_P = \frac{2(1-\sigma_Pa_P)}{T_R}.
\end{equation}
Then using Eq. \eqref{pumpdecayrate} in Eq. \eqref{pumpstrengthinring1}, the pumping strength in the ring can be written as
\begin{align}
    \label{pumpinring}
    g(\tilde{t}\,) &= \frac{g_0}{2} \frac{\kappa_P a_P}{\sqrt{1-\sigma_P a_P}}\sqrt{\frac{\tilde{\tau}}{8\ln(2)}}\exp\left(\frac{-2\ln(2)\,\tilde{t}^2}{\tilde{\tau}^2}\right)\nonumber
    \\
    &\times \sqrt{\pi}{\rm e}^{z(\tilde{t}\,)^2}{\rm erfc}[z(\tilde{t}\,)].
\end{align}
This completes the parameterization of our model.  

Our focus now is to examine the numerical results for the correlation variance in Eq. \eqref{correlationfunc} found by solving the dynamic equations with the pump field given by the expression in Eq. \eqref{pumpinring}. To do this, we use a 4th-order Runge-Kutta method which has a run-time of about 1 millisecond on a standard PC. Our model depends on the parameters $g_0$, $\kappa_P$, $\sigma_P$, and $a_P$, and the difference in the loss rates $\zeta$. Thus, our results are effectively independent of the ring radius, pump amplitude, and the nonlinear parameter. However, to make connection to a specific, realistic system, we choose a ring of radius $R = 20 \,{\rm \mu m}$ made from AlGaAs, with an effective second-order nonlinear susceptibility of $\chi^{(2)} = 11\,{\rm pm/V}$ \cite{Yang2007EnhancedResonator}. Additionally, we let the central frequency and amplitude of the pump pulse be $\omega_P = 2\pi \times 128.9\,{\rm THz}$  ($\lambda_P = 775\, {\rm nm}$) and $\mathcal{E}_0 = 1\,{\rm MV/cm}$. The frequencies of the signal and idler fields are  $\omega_1 = 2\pi \times 64.68\,{\rm THz}$ ($\lambda_1 = 1545\, {\rm nm}$) and $\omega_2 = 2\pi \times 64.27\,{\rm THz}$  ($\lambda_2 = 1555\, {\rm nm}$), respectively. With these parameter choices, $g_0 = 4$.

In our previous work \cite{Vendromin2020OptimizationStates}, we derived the optimum value for the pulse duration $\tilde{\tau}_{opt}$ by taking the derivative of the pumping strength in Eq. \eqref{pumpinring} with respect to $\tau$, at the peak value, and set it equal to zero. This results in the following accurate estimate of the optimum pulse duration of
\begin{equation}
    \label{optimumtau}
    \tilde{\tau}_{opt} = 0.684\sqrt{8\ln(2)}.
\end{equation}
The prefactor of 0.684 in this expression arises from the solution of a transcendental equation, and this can be evaluated to arbitrary accuracy. The optimum value for $\sigma_P$ can be obtained \cite{Vendromin2020OptimizationStates} by taking the derivative of the pumping strength with respect to $\sigma_P$, at the peak value and  $\tilde{\tau}=\tilde{\tau}_{opt}$, and set it equal to zero. For a loss parameter of $a_P = 0.99$, this results in an optimum value of $\sigma_P = 0.868$. 

\begin{figure}[htbp]
        \centering
        \begin{subfigure}[b]{.45\textwidth}
            \centering
            \includegraphics[width=\textwidth]{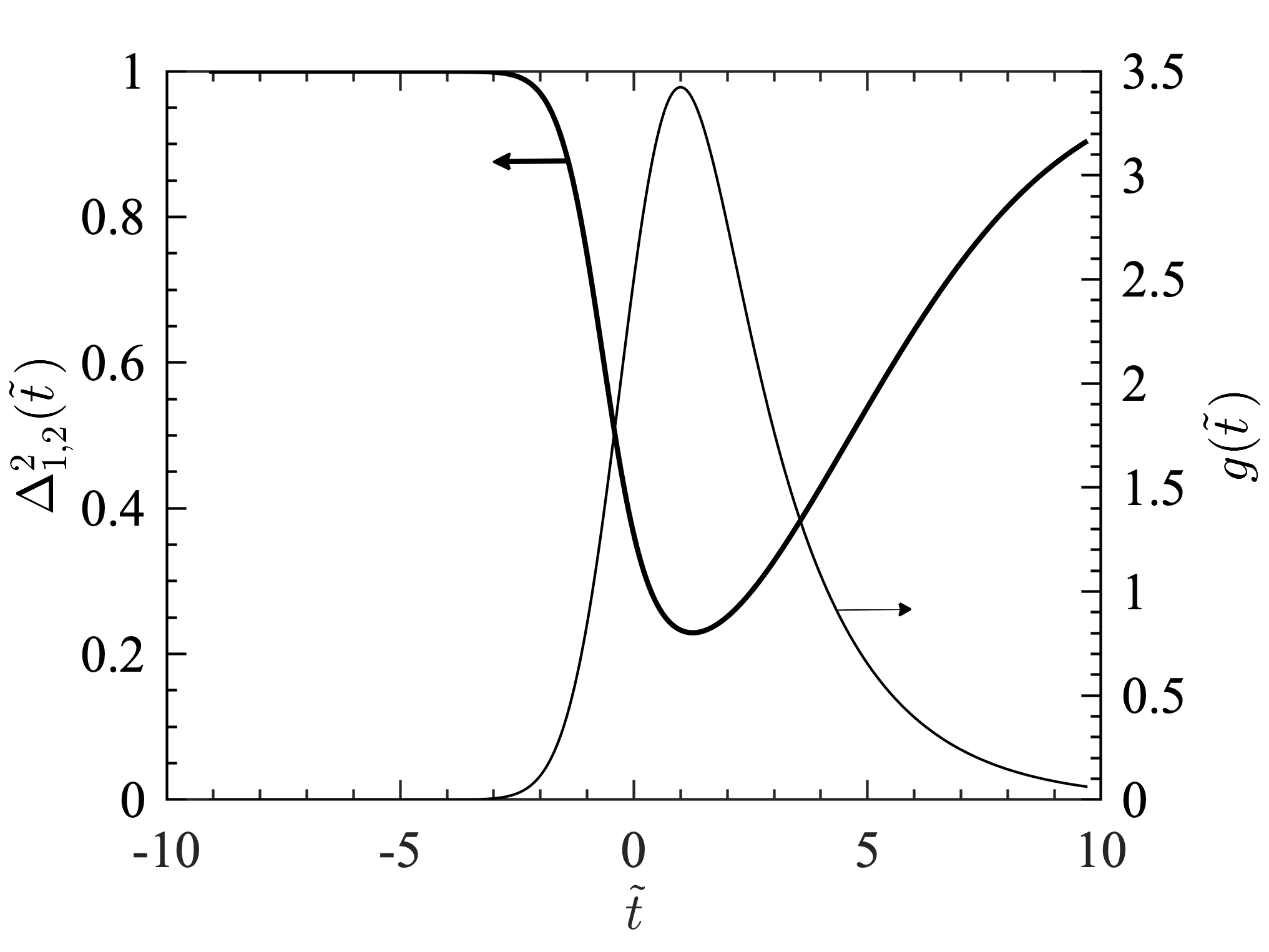}
            \caption[]%
            
            \label{fig:entanglemnt_ap=0.99_z=0}
        \end{subfigure}
        \vfill
        \begin{subfigure}[b]{.45\textwidth}  
            \centering 
            \includegraphics[width=\textwidth]{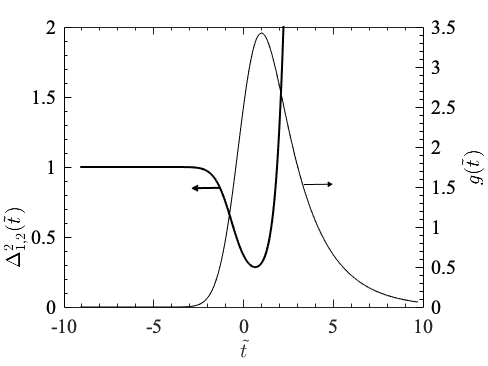}
            \caption[]%
            
            \label{fig:entanglemnt_ap=0.99}
        \end{subfigure}
      \caption{The time-dependent correlation variance for \small (a) $\zeta = 0$, and (b) $\zeta = 1/3$, pumping strength $g_0=4$, using a pulsed pump of duration $\tilde{\tau}_{opt}$, a pump scattering loss of $a_P=0.99$, and optimum self-coupling constant $\sigma_P = 0.868$. Also shown is the time-dependent pumping strength.}
\label{fig:entangledynamics}
\end{figure}

In Fig. \ref{fig:entanglemnt_ap=0.99_z=0} we plot the correlation variance when the decay rates of the two modes are set to be the same by letting $\zeta = 0$. Since the photons in modes 1 and 2 are scattered from the ring at the same rate, the photon pairs are not separated due to scattering and the state in the ring remains an entangled state with $\Delta^2_{1,2} < 1$. 

In Fig. \ref{fig:entanglemnt_ap=0.99} we again plot the correlation variance, but let $\zeta = 1/3$, which gives $\Gamma_1 = 2 \Gamma_2$ and $\Gamma_2 = 2\Gamma_P/3$. The discrepancy between the two decay rates means that photons are being scattered from mode 1 twice the rate as from mode 2. Effectively, this means that half of the generated entangled state is being traced-out by the scattering; $i.e.$, photon 1 escapes the ring while photon 2 stays inside the ring. Thus, the state in the ring evolves into a thermal state. This is why the correlation variance in Fig. \ref{fig:entanglemnt_ap=0.99} goes above 1 after $\tilde{t}\approx 2$. After this time, the correlation variance becomes as large as 60 (not shown in plot) and the state in the ring is essentially a separable two-mode thermal state.

\begin{figure}
    \centering
    \includegraphics[scale=1]{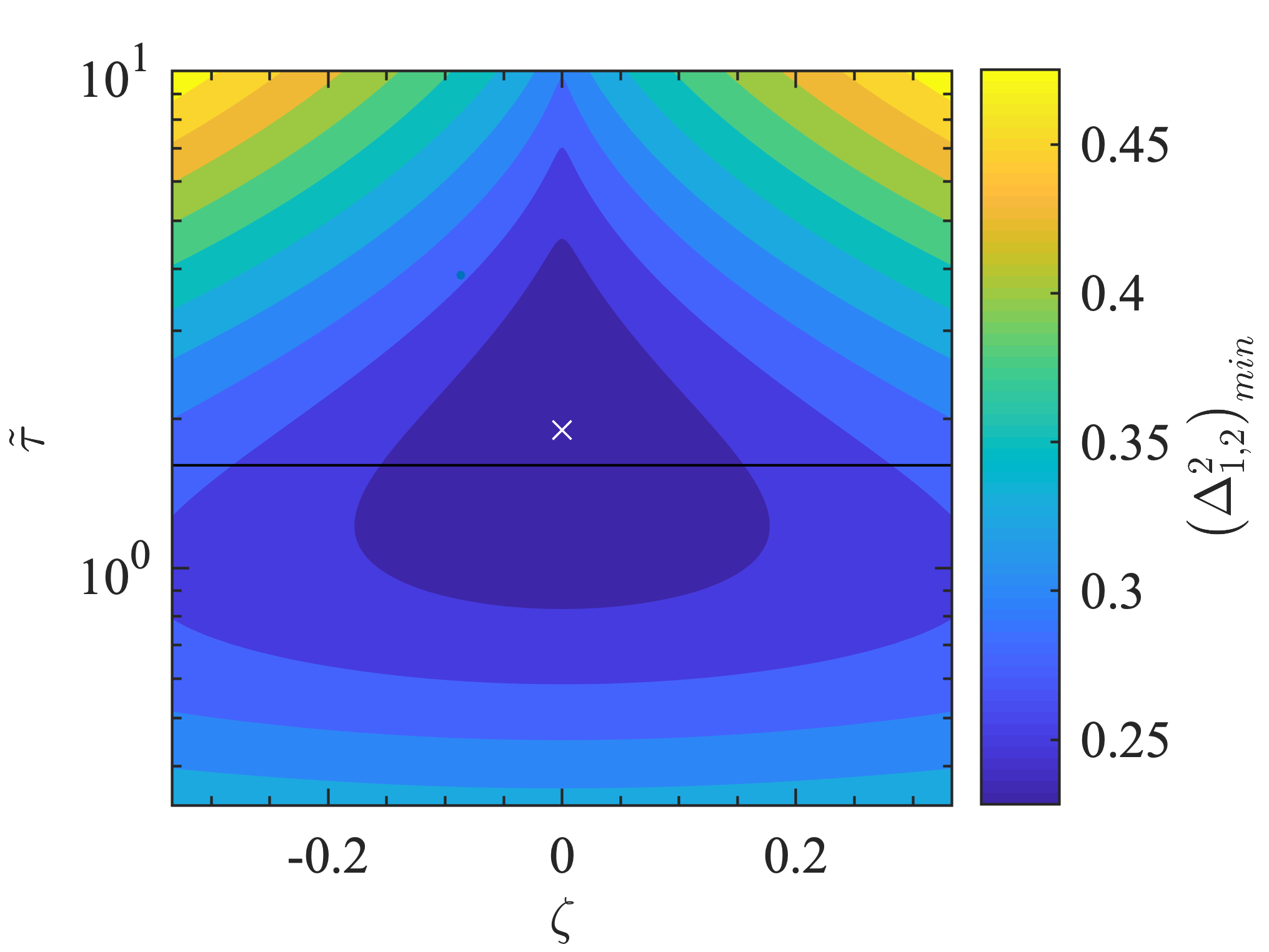}
    \caption{The minimum value of the correlation variance $(\Delta_{1,2}^2)_{min}$ as a function of the difference between the losses of the two-modes $\zeta$ and the pump pulse duration $\tilde{\tau}$. The optimum pulse duration $\tilde{\tau}_{opt}$ given by Eq. \eqref{optimumtau} is shown with the black line, and the global minimum is indicated by the white cross. All other parameters are the same as in Fig. \eqref{fig:entangledynamics}.}
    \label{fig:min3d}
\end{figure}

In Fig. \ref{fig:min3d} the minimum of the correlation variance, $\Delta^2_{1,2}(t_{min})$,  is shown as a function of $\tilde{\tau}$ and $\zeta$.  The first thing to notice is that it is symmetric about $\zeta = 0$. This is because the product $\zeta(n_2 - n_1)$ in Eq. \eqref{max_entangle} is always positive. Next we observe that for a given $\tau$, the best entanglement is always where $\zeta = 0$, because this makes the numerator of Eq. \eqref{max_entangle} as small as possible (\textit{i.e.}, equal to 1).  The global minimum in $\Delta^2_{1,2}(t_{min})$ (indicated by the white cross) occurs with the $\tau$ that makes $g(t_{min})$ as large as possible (see Eq. \eqref{approx_max_entangle}), which occurs approximately at $\tilde{\tau}_{opt}$ (black line) given in Eq. \eqref{optimumtau}. When we increase or decrease the pulse duration from the optimum value, it causes $\Delta^2_{1,2}(t_{min})$ to increase. This is because the energy of our pulse is independent of its duration, so for long pulses its amplitude scales as $1/\sqrt{\tau}$ and for short pulses the energy is injected over too short a time interval for the intensity to buildup. In both cases, for a long or short pulse, the pumping strength decreases and there is less entanglement. When $|\zeta|>0$, the entangled state is degraded by the unequal scattering of its photon pairs, and by increasing $\tau$ we observe a rapid decrease of the entanglement. Conversely, when $\zeta = 0$ and $\tau$ is increased, then the decrease of the entanglement is less rapid, because both photons are scattered at the same rate and the entanglement is better. However, when the pulse duration is shorter than the optimum,  the entanglement is not very sensitive to $|\zeta|$. 

\begin{figure}[htbp]
        \centering
            \includegraphics[scale=1]{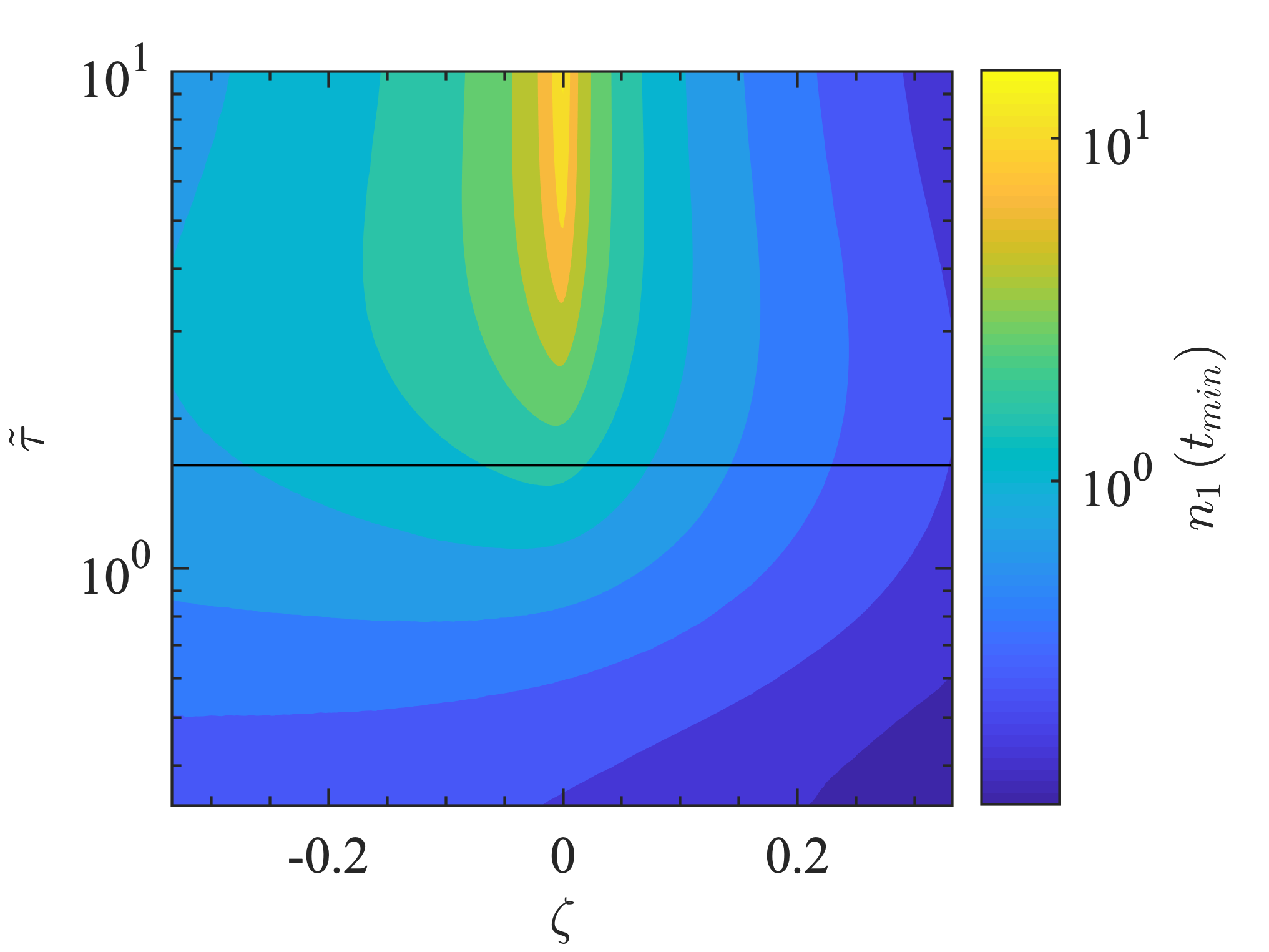}
      \caption{\small The average number of thermal photons in mode 1, evaluated at the time when the correlation variance is minimum, $n_1(t_{min})$, as a function of $\zeta$ and $\tilde{\tau}$. All parameters are the same as in Fig. \eqref{fig:min3d}}
      \label{fig:nth1}
\end{figure}

The thermal photon number in mode 1, evaluated at $t_{min}$, is shown in Fig. \ref{fig:nth1} as a function $\tilde{\tau}$ and $\zeta$ for the same parameters as above. A relatively large thermal photon number implies that there is more generation of squeezed light, but if the thermal photon number becomes too large, it destroys the entanglement (see Eq. \eqref{correlationfunc}). There is an average thermal photon number of approximately $2$ for each mode at the global minimum of the correlation variance. Where the average thermal photon number is $~10$ there is more generation of squeezed light, but the entanglement is weaker due to the increased thermal noise. The thermal photon number in mode 2 is a mirror reflection, about $\zeta = 0$, of mode 1, which can be shown by letting $\zeta \rightarrow -\zeta$ in Eqs. \eqref{nth1_deq_final} and \eqref{nth2_deq_final}.

\begin{figure}
    \centering
    \includegraphics{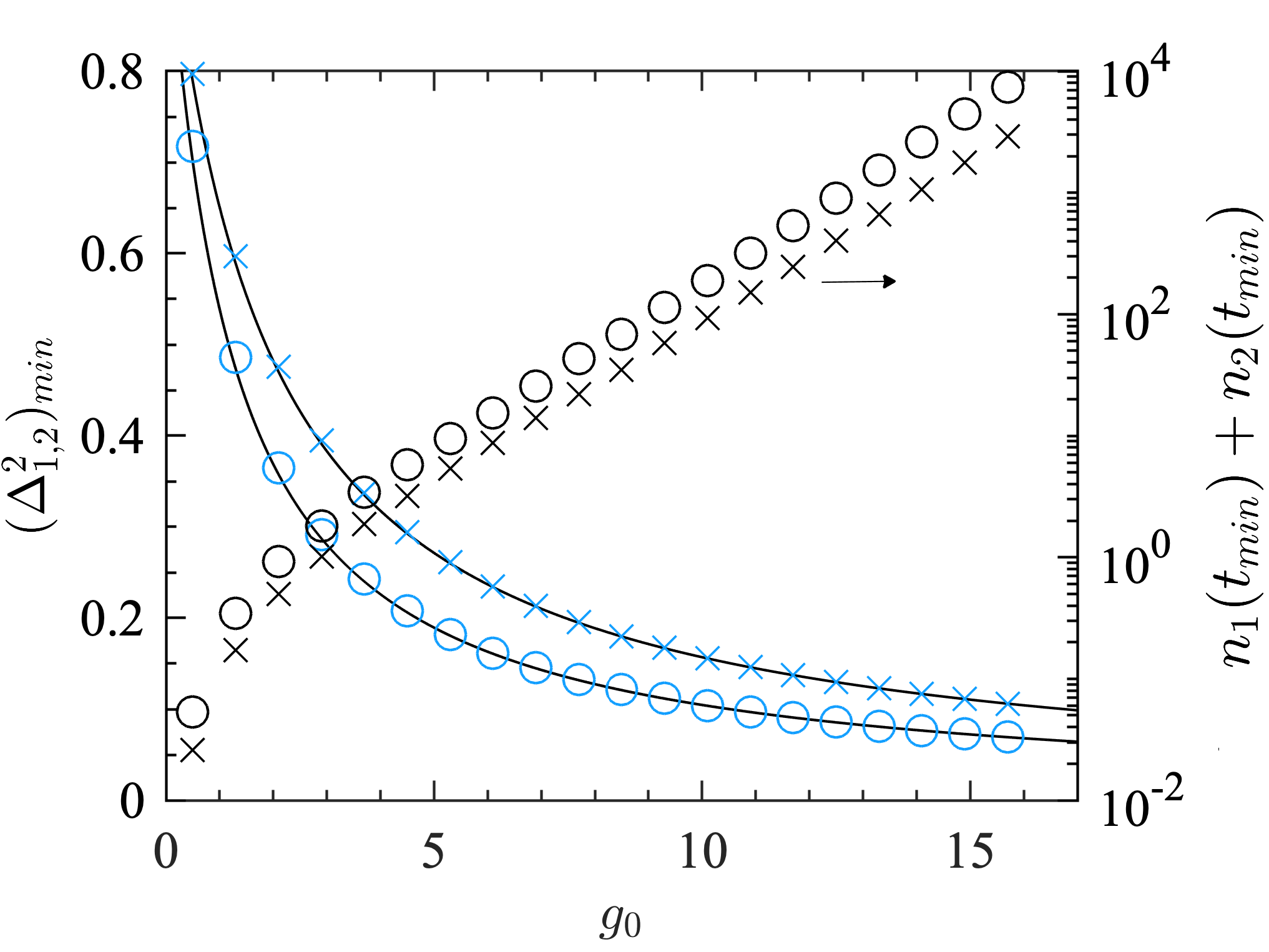}
    \caption{The global minimum in the correlation variance as a function of $g_0$ for pump scattering loss and optimum self-coupling constant of $a_P = 0.99$ and $\sigma_P = 0.868$ (circle) and $a_P = 0.75$ and $\sigma_P = 0.451$ (cross), respectively. Also shown is the total number of thermal photons evaluated at the global minimum for each of these losses.}
    \label{fig:minvsg0}
\end{figure}

In both Fig. \ref{fig:entanglemnt_ap=0.99_z=0} and Fig. \ref{fig:entanglemnt_ap=0.99}, the minimum value of the correlation variance occurs very close to the time when the pumping strength is maximum in the ring ($g_{max}$). Thus, in  Eq. \eqref{approx_max_entangle}, which gives the approximate minimum value of the correlation variance, we can let $g(t_{min})=g_{max}$. In our previous work \cite{Vendromin2020OptimizationStates} we showed that this is a valid approximation when the input pulse duration is longer than $T_R$. Using Eq. \eqref{pumpinring} it can be shown that the pumping strength peaks at $\tilde{t} = 1$ for the optimum pulse duration. Putting $\tilde{t} = 1$ and $\tilde{\tau}_{opt}$ into the pumping strength in Eq. \eqref{pumpinring} gives the following expression for $g_{max}$:
\begin{equation}
    \label{gmax}
    g_{max} = 0.653 g_0\frac{\kappa_P a_P}{\sqrt{1-\sigma_P a_P}}.
\end{equation}
Therefore, we can use Eq. \eqref{gmax} in Eq. \eqref{approx_max_entangle} in place of $g(t_{min})$ to obtain an expression for the minimum of the correlation variance when $|\zeta|\ll 1$ in terms of the pump parameters $g_0$, $\kappa_P$, $\sigma_P$, and $a_P$ only,
\begin{equation}
\label{mincorrelation}
    \left(\Delta^2_{1,2}\right)_{min} = \left[1+0.653 g_0\frac{\kappa_P a_P}{\sqrt{1-\sigma_P a_P}}\right]^{-1}.
\end{equation}
The optimum coupling constant $\sigma_P$ that minimizes Eq. \eqref{mincorrelation} for a fixed $g_0$ and $a_P$ is,
\begin{equation}
    \label{sigmaopt}
    \sigma_P = \frac{1-\sqrt{1-a_P^2}}{a_P}.
\end{equation}
Putting this optimum coupling constant into Eq. \eqref{mincorrelation} gives a good approximation to the global minimum in the correlation variance for a given $g_0$ and loss $a_P$. In Fig. \ref{fig:minvsg0}, the global minimum in the correlation variance is shown as a function of $g_0$ for $a_P = 0.99$ (circle) and $a_P = 0.75$ (cross). The solid lines are the expression in Eq. \eqref{mincorrelation}, and they show excellent agreement with the full numerical simulations across a wide range of $g_0$ values for the two scattering losses. Also shown is the sum of the thermal photon numbers in the two modes. This shows that the cost associated with decreasing the correlation variance is an increase in the overall thermal noise in the ring. By increasing the scattering loss we observe a decrease in the entanglement and the total thermal photon number. In order to recover the entanglement we can increase $g_0$, but this causes an increase in the thermal noise. For example, when $a_P =0.99$ and $g_0 = 4$, the minimum of the correlation variance is about $0.25$ and there are a total of about $4$ thermal photons. If we increase the scattering loss to $a_P =0.75$ and keep $g_0 =4$, then the variance jumps to about $0.35$ and the total thermal photon number decreases to about $2$. In order to recover the variance, we would need to increase to $g_0 = 7$, and as a result we would increase the total thermal photon number to about $20$. 

\begin{figure*}
    \centering
    \begin{subfigure}{0.32\textwidth}
    \includegraphics[width=\textwidth]{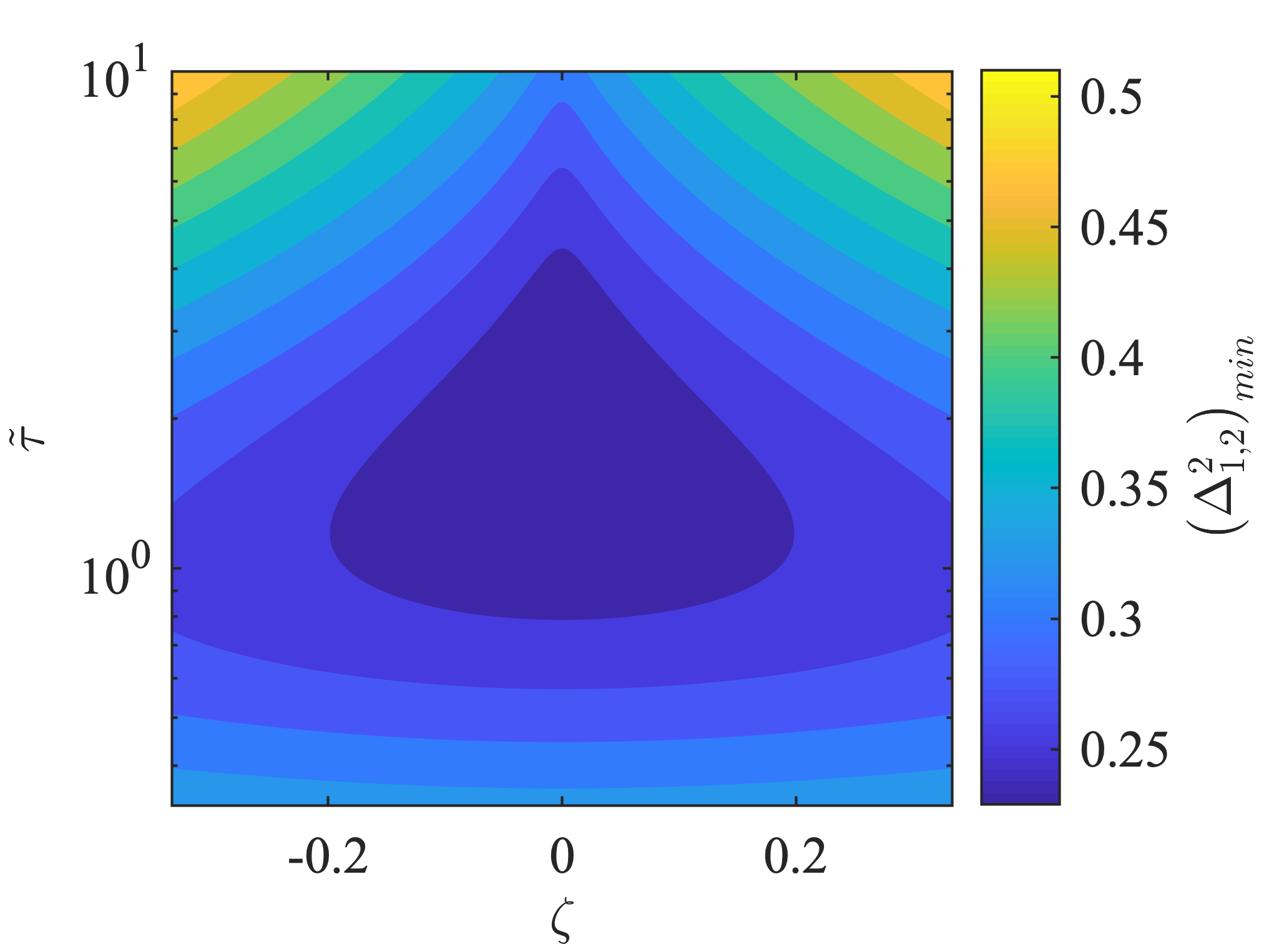}
    \caption{}
    \label{5mrad}
    \end{subfigure}
    \begin{subfigure}{0.32\textwidth}
    \includegraphics[width=\textwidth]{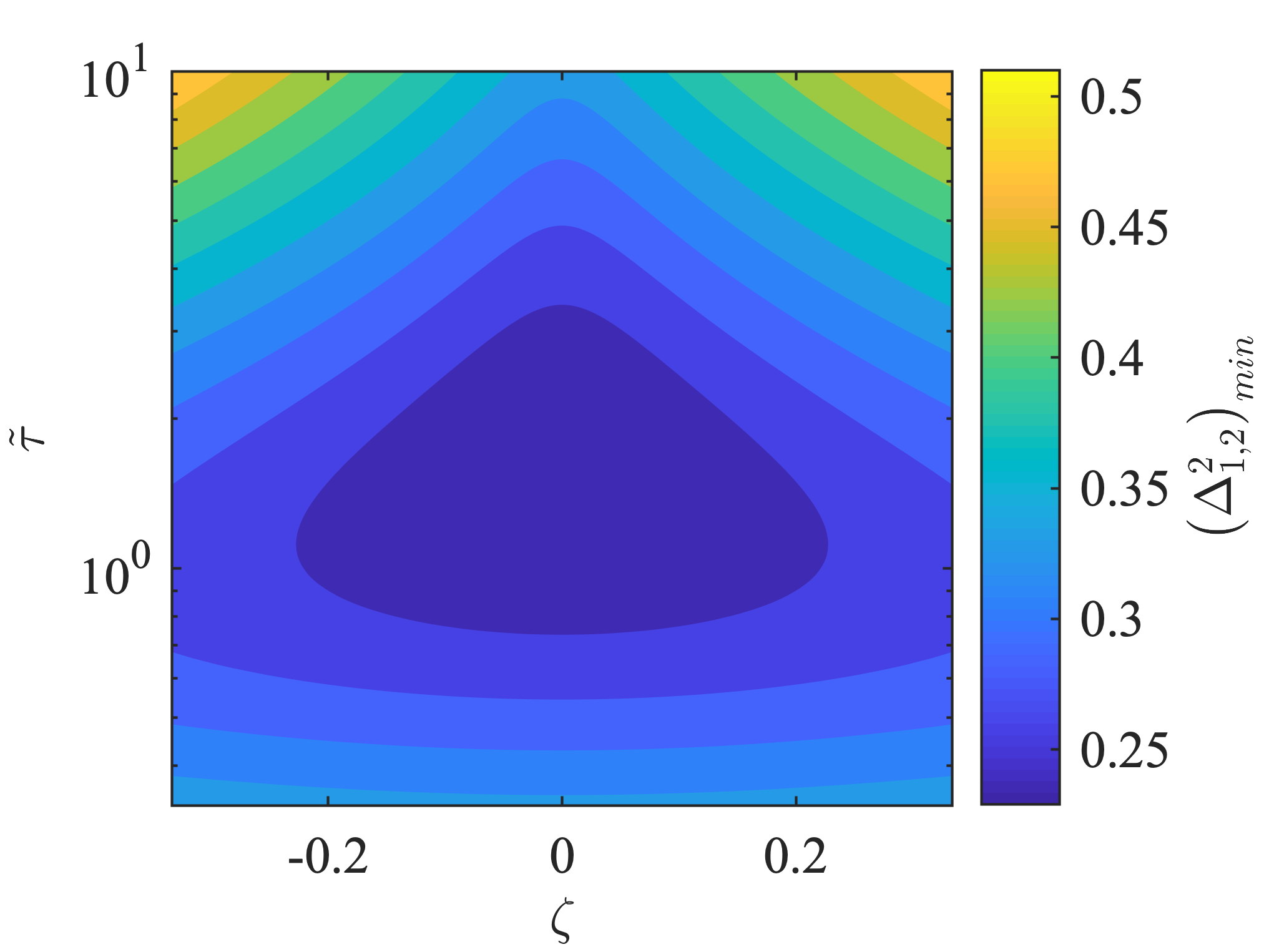}
    \caption{}
    \label{20mrad}
    \end{subfigure}
    \begin{subfigure}{0.32\textwidth}
    \includegraphics[width=\textwidth]{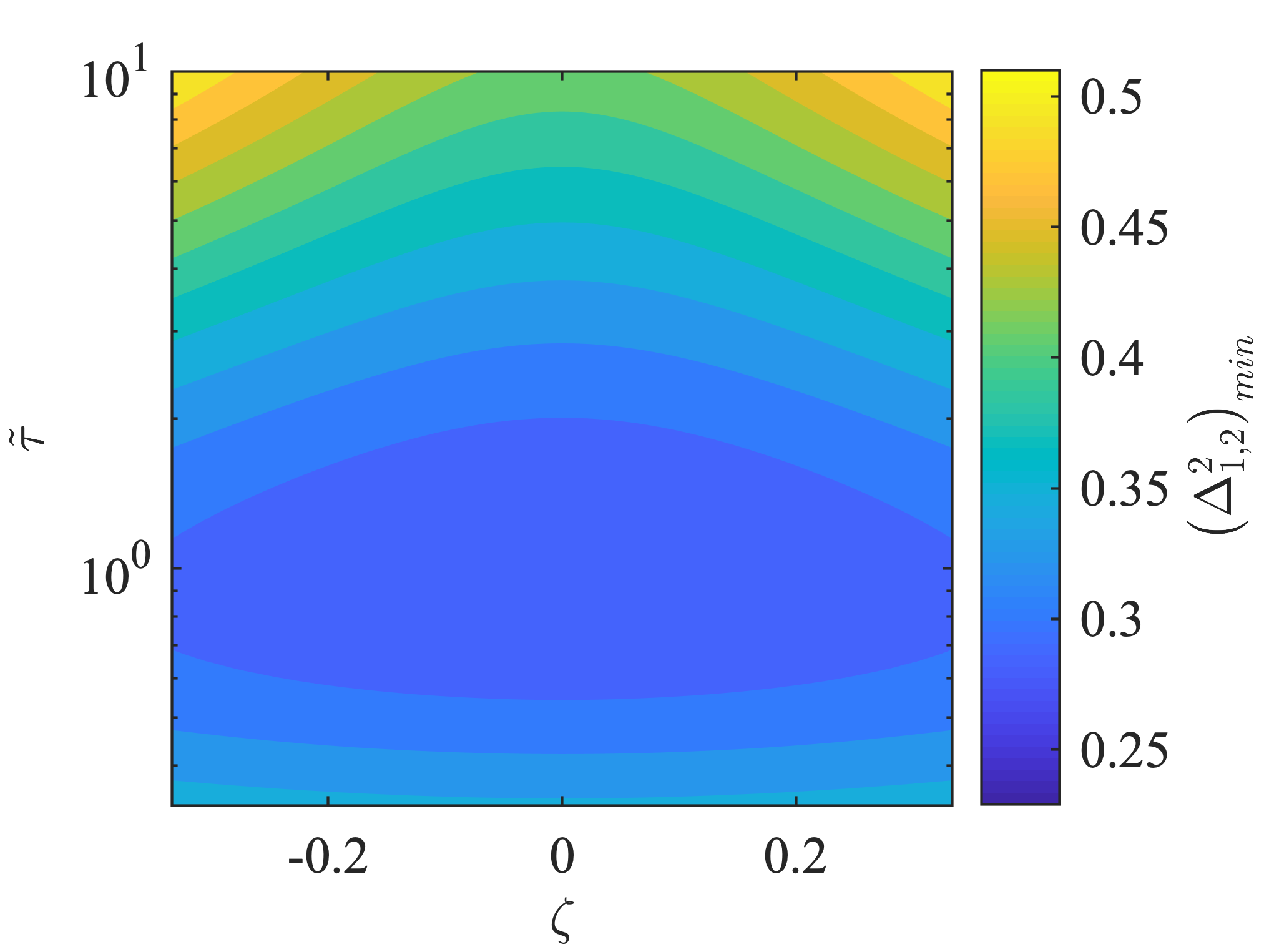}
    \caption{}
    \label{100mrad}
    \end{subfigure}
    \caption{The minimum in the correlation variance as a function fo $\tilde{\tau}$ and $\zeta$ for homodyning angles of (a) $5\, {\rm mrad}$, (b) 20\,{\rm mrad}, and (c) $100 \,{\rm mrad}$. All other parameters are the same as in Fig. \eqref{fig:min3d}. }
\end{figure*}

The expression for the correlation variance in Eq. \eqref{correlationfunc} is what one would infer from measurements in a homodyne detection experiment, if the local oscillators are perfectly phase matched to the squeezing phase $\phi(t)$. It what follows, we will allow for a small angle deviation $\delta \theta$ from perfect phase matching and see how this affects the correlation variance. The correlation variance when there is an angular deviation becomes
\begin{equation}
    \label{corr_deviated}
    \Delta_{1,2}^2 = \left(1+n_1+n_2\right)\left[\cosh(2u)-\cos(\delta\theta)\sinh(2u)\right].
\end{equation}
If $\delta \theta = 0$ we recover the perfect case in Eq. \eqref{correlationfunc}. In Fig. \ref{5mrad} we show the minimum of the correlation variance for an angular deviation of $5\,{\rm mrad}$, with $a_P = 0.99$, $\sigma_P =0.868$, and $g_0 = 4$. This angular deviation was measured in a recent experiment \cite{Shi2018DetectionAbsorption}. Overall, the correlation variance is not changed significantly from the ideal case (when there is no offset), but there is a small increase in the global minimum from approximately 0.228 when there is no offset to approximately 0.229. In Fig. \ref{20mrad} and Fig. \ref{100mrad} the angle offset is increased to $\delta \theta = 20 \,{\rm mrad} $ and $\delta \theta = 100\,{\rm mrad}$, respectively, while keeping all other parameters the same. For these two offsets, the global minimum increases to approximately 0.235 and 0.282, respectively. The general trend created by increasing the angle offset is that the global minimum of the correlation variance shifts to shorter pulse durations and the optimal region flattens out; becoming more sensitive to pulse duration and less sensitive to $\zeta$. Thus, the discrepancy between the two decay rates does not destroy the entanglement as much when there is an angle offset. Instead, the pulse duration is much more destructive to the entanglement. When there is an angle offset, it is always better to shorten the optimum pulse duration to improve entanglement. 

\section{Conclusion}
\label{conclusion}
In conclusion, we have shown that the exact solution to the Lindblad master equation for a two-mode cavity pumped with a classical optical pulse is a two-mode squeezed thermal state for all time. We derived a semi-analytic solution for the time-dependent squeezing amplitude, and thermal photon numbers for the two modes, which we then used to derive an expression for the maximum entanglement in the cavity (see Eq. \eqref{max_entangle}). The main parameter that affects the amount of entanglement is the difference in the cavity losses between the two modes $\zeta$. In the ideal case, the losses for the two modes are the same $\zeta = 0$ and the correlation variance just scales inversely proportional to the peak pumping strength in the cavity (see Eq. \eqref{approx_max_entangle}).  However, the trade-off is that when the pumping strength is increased the total thermal noise also increases. 

We applied our theory to the important example of generating CV entanglement in a ring resonator with different losses for each mode. We derived a semi-analytic expression for the maximum entanglement in the ring (see Eq. \eqref{mincorrelation}) that depends only on the pump scattering loss and coupling parameters. If the two modes have unequal losses we demonstrated that it is always better to decrease the pulse duration from the optimum in order to achieve better entanglement. We considered the case of a phase offset in a homodyne detection experiment that would cause a degradation of the measured entanglement, and we showed that the maximum entanglement could be recovered to a high degree by reducing the pulse duration from the optimum. Additionally, when the phase offset is increased, it was shown that the entanglement is less sensitive to the unequal losses of the two modes.

These results will be of use to researchers that are trying to optimize CV entanglement in lossy cavities when the losses of each mode are different. In future work, we will apply this theory to the generation of two-mode squeezed light in a slab photonic crystal, where a three-mode cavity is side-coupled to a defect waveguide. 
\section*{Acknowledgements}
This work was supported by Queen’s University and the Natural Sciences and Engineering Research Council of Canada (NSERC).

\bibliography{apssamp}

\begin{thebibliography}{22}
\bibitem [{ini()}]{initial_phase_note}%
  \BibitemOpen
  \href@noop {} {}\bibinfo {note} {Although the phase with a positive sign in
  front of the $\pi/2$ term is also a solution, that solution gives
  $du(t_i)/dt<0$, which results in a negative $u$, which is not allowed, since
  by definition $u$ is a positive amplitude.}\BibitemShut {Stop}%
\end{thebibliography}

\appendix
\section{Simplifying Eq. \eqref{timederivative2_terms}}
\label{appndx1}
In this section we simplify the RHS of Eq. \eqref{timederivative2_terms} in order to find a condition for when the equality is true. We define the first two terms on the RHS of \eqref{timederivative2_terms} to be
 \begin{equation}
     T1 =\frac{d\hat{\rho}_{th}^{-1/2}}{dt}\hat{S}^\dagger\hat{\rho}\hat{S}\hat{\rho}_{th}^{-1/2} +\hat{\rho}_{th}^{-1/2}\hat{S}^\dagger\hat{\rho}\hat{S}\frac{d\hat{\rho}_{th}^{-1/2}}{dt},
     \label{firstterm}
 \end{equation}
 and then using Eq. \eqref{constantofmotion2} to simplify, we obtain
\begin{align} \label{firstterm_final}
  T1&= 2 \frac{d\hat{\rho}_{th}^{-1/2}}{dt}\hat{\rho}_{th}^{1/2}\nonumber
  \\
  &= \sum_{j=1}^2\left(-\hat{n}_j \frac{\dot{x}_j}{x_j} + \hat{\mathbbm{1}}\frac{\dot{x}_j}{1-x_j}\right).
\end{align}
We used the two-mode thermal state in Eq. \eqref{2modethermalstate} to get the last line.


We define the middle two terms on the RHS of \eqref{timederivative2_terms} to be
 \begin{equation}
     T2 =\hat{\rho}_{th}^{-1/2}\frac{d\hat{S}^\dagger}{dt}\hat{\rho}\hat{S}\hat{\rho}_{th}^{-1/2} + \hat{\rho}_{th}^{-1/2}\hat{S}^\dagger\hat{\rho}\frac{d\hat{S}}{dt}\hat{\rho}_{th}^{-1/2},
     \label{secondterm}
 \end{equation}
 and then using Eq. \eqref{constantofmotion2} to simplify, we obtain
\begin{align}\label{secondterm_2}
  T2&= \hat{\rho}_{th}^{-1/2}\frac{d\hat{S}^\dagger}{dt}\hat{S}\hat{\rho}_{th}^{1/2} - \hat{\rho}_{th}^{1/2}\frac{d\hat{S}^\dagger}{dt}\hat{S}\hat{\rho}_{th}^{-1/2}.
\end{align}
To simplify this expression we will need to take the time derivative of the two-mode squeezing operator in Eq. \eqref{squeezingop}, which is not straightforward, and thus requires us to be careful. We let $\hat{S} = \exp(\sigma)$, where $\sigma =\xi^*(t)\hat{b}_1\hat{b}_2 - h.c. $ and  $\hat{S}^\dagger = \exp(-\sigma)$, then
\begin{align}
\label{Sderiv_1}
     \frac{d\hat{S}^\dagger}{dt} &= \frac{d}{dt}\left( 1 - \sigma + \frac{\sigma^2}{2!} - \frac{\sigma^3}{3!} + ...\right) \nonumber
     \\
     &= \sum_{n=0}^{\infty} \sum_{k=0}^{\infty}(-1)^{n+k+1}\frac{\sigma^n \dot{\sigma}\sigma^{k}}{(n+k+1)!}\nonumber
    \\
    &= -\int_{0}^{1}d\lambda \exp(-\lambda \sigma) \dot{\sigma}\exp(\lambda \sigma)\exp(-\sigma), 
\end{align}
where the integral in the last line can be shown to be equivalent to the sum on the previous line by expanding the exponential operators in a power series in $\sigma$ and doing the integration over $\lambda$ from 0 to 1. Multiplying Eq. \eqref{Sderiv_1} by $\hat{S}$ from the right, we obtain
\begin{equation}
\label{Sderiv_2}
    \frac{d\hat{S}^\dagger}{dt}\hat{S}  = -\int_{0}^{1}d\lambda \exp(-\lambda \sigma) \dot{\sigma}\exp(\lambda \sigma). 
\end{equation}
Using the well-known Baker-Campbell-Hausdorff formulae on the integrand of Eq. \eqref{Sderiv_2}, and then integrating over $\lambda$ in each term in the series, we obtain
\begin{align}
\label{Sderiv_3}
    \frac{d\hat{S}^\dagger}{dt}\hat{S}  &= -\dot{\sigma} + \frac{1}{2!}[\sigma,\dot{\sigma}] -\frac{1}{3!}[\sigma,[\sigma,\dot{\sigma}]] + ...\nonumber
    \\
    &=\sum_{n=1}^{\infty} (-1)^n \frac{L^{(n)}}{n!},
\end{align}
where the first three terms in $L^{(n)}$ are defined as $L^{(1)}\equiv \dot{\sigma}$, $L^{(2)}\equiv [\sigma,\dot{\sigma}]$, and $L^{(3)}\equiv [\sigma,[\sigma,\dot{\sigma}]]$. In general, we can write $L^{(n)} = [\sigma,L^{(n-1)}]$ for $n\ge2$. It is straightforward to show that,
\begin{align}
    \label{Sderiv_4}
    L^{(1)} &= \dot{u}\left({\rm e}^{-i\phi}\hat{b}_1\hat{b}_2 - {\rm e}^{i\phi}\hat{b}_1^\dagger \hat{b}_2^\dagger\right) \nonumber
    \\
    &- iu\dot{\phi}\left({\rm e}^{-i\phi}\hat{b}_1\hat{b}_2  + {\rm e}^{i\phi}\hat{b}_1^\dagger \hat{b}_2^\dagger\right)
    \\
     L^{(n)} &= -\frac{i\dot{\phi}}{2}\left(\hat{n}_1 + \hat{n}_2  + 1 \right)(2u)^n,\,\, \text{even}\, n\ge 2 \label{Sderiv_5}
     \\
     L^{(n)} &= -\frac{i\dot{\phi}}{2}\left({\rm e}^{-i\phi}\hat{b}_1\hat{b}_2  + {\rm e}^{i\phi}\hat{b}_1^\dagger \hat{b}_2^\dagger\right)(2u)^n,\,\, \text{odd}\, n\ge 3 \label{Sderiv_6}.
\end{align}
We then use Eqs. \eqref{Sderiv_4} to \eqref{Sderiv_6} in Eq. \eqref{Sderiv_3} to simplify the derivative. The sum over even $n$ converges to $\cosh(2u)-1$  and  the sum over odd $n$ converges to $\sinh(2u)-2u$.  Using these results, we put this simplified form of Eq. \eqref{Sderiv_3} into Eq. \eqref{secondterm_2}, along with the two-mode thermal state, to obtain
\begin{align}
\label{secondterm_final}
  T2&= \frac{1-x_1x_2}{\sqrt{x_1x_2}}\left(\dot{u}\hat{U} + \frac{1}{2}\sinh(2u)\dot{\phi}\hat{V}\right),
\end{align}
where $\hat{U} = \hat{b}_1\hat{b}_2\exp(-i\phi)+ h.c.$ and $\hat{V} =-i\hat{b}_1\hat{b}_2\exp(-i\phi)+ h.c.$. We define the last term of Eq. \eqref{timederivative2_terms} to be
\begin{equation}
    T3 = \hat{\rho}_{th}^{-1/2}\hat{S}^\dagger\frac{d\hat{\rho}}{dt}\hat{S}\hat{\rho}_{th}^{-1/2}.
    \label{thirdterm}
\end{equation}
Using Eq. \eqref{lindblad} in Eq. \eqref{thirdterm} and simplifying by using Eq. \eqref{constantofmotion2} gives 
\begin{align}\label{thirdterm_2}
    T3 &= -\frac{i}{\hbar}\left( \hat{\rho}_{th}^{-1/2}\hat{S}^\dagger \hat{H}\hat{S}\hat{\rho}_{th}^{1/2}  - \hat{\rho}_{th}^{1/2}\hat{S}^\dagger \hat{H}\hat{S}\hat{\rho}_{th}^{-1/2} \right) + \nonumber
    \\
    &+\sum_{j=1}^2\Gamma_j \hat{\rho}_{th}^{-1/2}\hat{S}^\dagger \hat{b}_j\hat{S}\hat{\rho}_{th}^{1/2}\hat{\rho}_{th}^{1/2}\hat{S}^\dagger \hat{b}^\dagger_j\hat{S}\hat{\rho}_{th}^{-1/2} - \nonumber
    \\
    &-\frac{1}{2}\sum_{j=1}^2\Gamma_j\left( \hat{\rho}_{th}^{-1/2}\hat{S}^\dagger \hat{n}_j\hat{S}\hat{\rho}_{th}^{1/2} +\hat{\rho}_{th}^{1/2}\hat{S}^\dagger \hat{n}_j\hat{S} \hat{\rho}_{th}^{-1/2} \right) 
\end{align}
 which can be simplified using the well-known Baker-Campbell-Hausdorff formulae to obtain
\begin{widetext}
\begin{align}\label{thirdterm_final}
    T3 &= \frac{1-x_1x_2}{\sqrt{x_1x_2}}\left[\frac{\omega_1+\omega_2}{2}\sinh(2u)\hat{V} +\left(\frac{i}{\hbar}\left(\mathcal{E}_P \gamma {\rm e}^{-i\phi}\cosh^2{u}+\mathcal{E}^*_P\gamma^* {\rm e}^{i\phi}\sinh^2{u}\right)\frac{\hat{U}+i\hat{V}}{2} + h.c.\right) \right] + \nonumber
    \\
    &+\left[\left(\frac{1+x_1x_2}{2\sqrt{x_1x_2}} -\sqrt{\frac{x_1}{x_2}} \,\right)\Gamma_1 + \left(\frac{1+x_1x_2}{2\sqrt{x_1x_2}} -\sqrt{\frac{x_2}{x_1}}\, \right)\Gamma_2  \right]\frac{1}{2}\sinh(2u)\hat{U} \nonumber
    \\
    &+\left[\Gamma_1(x_1-1)\cosh^2u + \left(x_1^{-1}-1\right)\Gamma_2\sinh^2u\right]\hat{n}_1+\left[\Gamma_2(x_2-1)\cosh^2u + \left(x_2^{-1} - 1\right)\Gamma_1\sinh^2u\right]\hat{n}_2 + \nonumber
    \\
    &+\left[\Gamma_1\left(x_1\cosh^2u - \sinh^2u\right) +\Gamma_2\left(x_2\cosh^2u - \sinh^2u\right) \right]\hat{\mathbbm{1}}.
\end{align}
\end{widetext}
Using expressions for $T1$, $T2$, and $T3$ (see Eqs. \eqref{firstterm_final}, \eqref{secondterm_final}, and \eqref{thirdterm_final}) in Eq. \eqref{timederivative2_terms}, formally we can write
\begin{equation}\label{characteristiceq2}
  0 = T1+T2+T3.  
\end{equation}
In order for the equality in Eq. \eqref{characteristiceq2} to be true for all times we must have that the coefficients in front of the operators $\hat{\mathbbm{1}}$, $\hat{n}_1$, $\hat{n}_2$, $\hat{U}$, and $\hat{V}$ are equal to zero for all times. Setting these equal to zero and solving, we obtain the coupled differential equations Eqs. \eqref{nth1_deq} to \eqref{phi_deq}.
\end{document}